\documentstyle [12pt] {article}
\title{A QUANTUM GATE AS A PHYSICAL  MODEL OF AN UNIVERSAL
ARITHMETICAL ALGORITHM WITHOUT CHURCH'S  UNDECIDABILITY AND
G$\ddot{O}$DEL'S INCOMPLETENESS}
\author{Vladan Pankovi\'c, Milan
Predojevi\'c\\
Department of Physics,\\Faculty of Natural Sciences and
Mathematics,\\21000 Novi Sad,Trg Dositeja Obradovi\'ca 4.,Serbia
and Montenegro,
\\vladanp@gimnazija-indjija.edu.yu}

\date {}
\begin{document}
\maketitle

\vspace {1cm}


\vspace {1.5cm}

\begin {abstract}
In this work we define an universal arithmetical algorithm, by
means of the standard quantum mechanical formalism, called
universal qm-arithmetical algorithm. By universal qm-arithmetical
algorithm any decidable arithmetical formula (operation) can be
decided (realized, calculated. Arithmetic defined by universal
qm-arithmetical algorithm called qm-arithmetic one-to-one
corresponds  to decidable part of the usual arithmetic. We prove
that in the qm-arithmetic the undecidable arithmetical formulas
(operations) cannot exist (cannot be consistently defined). Or, we
prove that qm-arithmetic has no undecidable parts. In this way we
show that qm-arithmetic, that holds neither Church's
undecidability nor G$\ddot {o}$del's incompleteness, is decidable
and complete. Finally, we suggest that problems of the foundation
of the arithmetic, can be solved by qm-arithmetic.
\end {abstract}
\vspace {1.5cm}

\section {Introduction}

It is well-known that {\it usual} arithmetic (and similar
theories, i.e. "related systems") according to Church's
undecidability theorem [1] and G$\ddot {o}$del's incompleteness
theorem [2], [3] is {\it undecidable} and {\it incomplete}. Or, in
the usual arithmetic there are {\it decidable} formulas
(operations), any of them can be decided (realized, calculated) by
universal (arithmetical) algorithm (corresponding, according to
Church's thesis [1], to universal Turing's machine). But, also in
the usual arithmetic (and {\it any its extension with similar
structure}, called related system) there are {\it undecidable}
formulas (operations) that cannot be decided by universal
algorithm. Usual arithmetic considers too that the addition and
multiplication (of the natural or whole numbers) are not really
the elementary operations since both, by Peano's induction axiom
[2], [3], can be reduced in the elementary "the immediate
successor of" operation. Finally, usual arithmetic considers
implicitly that the universal Turing's machine or any physical
model of the universal algorithm works according to classical
mechanics.

In this work we shall {\it firstly} define especial quantum gates,
{\it qm-adder} and {\it qm-multiplier}. They represent physical
models, that works according to quantum mechanics (precisely, {\it
standard quantum mechanical formalism}) [4]-[9], of the algorithms
for addition and multiplication. According to standard quantum
mechanical formalism (or {\it mathematical theory of the Hilbert's
space of the unit norm vectors}), qm-adder and qm-multiplier
represent really the elementary algorithms since neither qm-adder
nor qm-multiplier can be reduced in "the immediate successor of"
operation. {\it Secondly}, we shall introduce such {\it induction
rule} according to which an universal arithmetical algorithm, {\it
universal qm-arithmetical algorithm}, can be defined. By
qm-universal algorithm {\it any } decidable arithmetical formula
(operation) can be decided. Arithmetic defined by universal
qm-arithmetical algorithm, called {\it qm-arithmetic }, {\it
one-to-one corresponds}  to {\it decidable part of the usual
arithmetic}. {\it Thirdly}, we shall prove that in the
qm-arithmetic the undecidable formulas (operations) {\it cannot
exist }(cannot be consistently defined). Or, we shall prove that
qm-arithmetic {\it has no} undecidable parts. In this way we shall
show that {\it qm-arithmetic is decidable and complete }. {\it
Finally}, on the basis of the mentioned  proofs, we shall suggest
that all problems of the foundation of the arithmetic [1]-[3] can
be simply solved by changing of the usual by qm-arithmetic.

\section {Quantum adder}

We shall define quantum gate for realization of the well-known
arithmetical operation, addition, +, (and difference , - ) of the
whole or natural numbers in an effectively finite time interval.
It will be called qm-adder.

Let {\it H} be the infinite-dimensional Hilbert's space of the
quantum states (vectors) of the unit norm of some quantum  system.

It will be pointed out, even if it is well-known, that constant
(unit) norm condition represents one of the most important
condition of the standard quantum mechanical formalism and
corresponding mathematical theory of the Hilbert's space with
constant (unit)  norm of the states (vectors) [4]-[7]. In other
words in given physical formalism or corresponding mathematical
theory  this condition can be broken neither explicitly nor
implicitly. Especially, this condition must be satisfied by
unitary symmetric quantum mechanical dynamical evolution on a
quantum super-system or sub-system (by "extension" of some
sub-systems in a super-system). Also, this condition must be
satisfied by  measurement on a super-system or sub-system (by
"reduction" of some super-system in  its sub-systems). For
example, states of two sub-systems must have the same unit norm as
well as the state of corresponding super-system. All this
indicates the following very important consequence. In the quantum
mechanics, or in the theory of the Hilbert's space with constant
(unit) norm of the vectors, there is none addition (difference) of
the vectors norms equivalent  to addition (difference) of the
whole or natural numbers.

Let ${\it B} = {…,|-n>,…,|-1>,|0>,|1>,…,|n>,…}$ be an especially
chosen, called {\it computational}, complete basis in ${\it H}$.

Let $A$ and $B$ be two quantum systems, first one and second one,
whose quantum states belong to ${\it B}_{A}$ and ${\it B}_{B}$
(both equivalent to ${\it B}$) from ${\it H}_{A}$ and ${\it
H}_{B}$ (both equivalent to ${\it H}$) respectively.

Let short-lived (which means that here time dependence of
corresponding states and operators will not be given explicitly)
quantum mechanical dynamical interaction between $A$ and $B$ or
quantum mechanical dynamical evolution on the quantum super-system
$A+B$ (that holds both $A$ and $B$ representing its sub-systems)
be given, in its integral form, by
\begin {equation}
\hat {U}_{+}|n>|m> = |n>|n+m>    \hspace{0.5cm}for \hspace{0.5cm}
\forall |n> \in B_A , \forall  |m> \in B_B
\end {equation}
Here $\hat {U}_{+}$ represents the unitary evolution operator
determined completely by (1), $|n>$ and $|m>$ on the left hand of
the (1)  - initial quantum mechanical dynamical states of the $A$
and $B$, and $|n>$ and $|n+m>$ on the right hand of the (1)  -
final quantum mechanical dynamical states of the $A$ and $B$. Of
course $\hat {U}_{+}$ acts over Hilbert's space of the $A+B$,
${\it H}_{A}\otimes {\it H}_{B}$, whose computational basis is
${\it B}_{A}\otimes {\it B}_{B}$ , where $\otimes $ represents the
tensorial product. More precisely $\hat {U}_{+}$ one-to-one maps
${\it B}_{A}\otimes {\it B}_{B}$ in the ${\it B}_{A}\otimes {\it
B}_{B}$. (In (1) as well as in all further text tensorial product
of the quantum states will not be given explicitly.)

Here we shall analyze only states from computational bases.
Superposition of the states from the computational bases that
unambiguously exists will be not considered. However, it is not
hard to see that for $ \forall  n, m, p, q \in W$ and any  real
numbers $c$, $d$, $e$, $f$  that satisfy conditions
$|c|^{2}+|d|^{2}=1$ and $|e|^{2}+|f|^{2}=1$, from the (1) it
follows $\hat {U}_{+}(c|n>+d|p>)(e|m>+f|q>) = ce |n>|n+m> + cf
|n>|n+q>+ de|p>|p+m>+df|p>|p+q>$. It means that given qm-adder can
work by arbitrary superposition of the states from the
computational bases, i.e. by "qubits".

Obviously, according to (1), there is one-to-one correspondence
between $A$ initial  dynamical state  $|n>$ and $B$ initial
dynamical state $|m>$ on the one hand and first and second
addition arguments $n$ and $m$  on the other hand,  for $\forall
n, m \in W$, where $W$ represents the set of all whole numbers. It
means that initial states of the $A$ and $B$ can represent the
qm-adder inputs. Also, according to (1), there is one-to-one
correspondence between $B$ final dynamical state $|n+m>$ and
addition result $n+m$ for $\forall  n,m \in W$. It means that $B$
final state can represent the qm-adder output.

So, it is proved that described $A+B$ super-system with quantum
mechanical dynamics (1) represents a qm-adder. It can be added
that many quantum mechanical super-systems with two sub-systems
holds quantum mechanical dynamic equivalent or very similar to
(1). For example quantum mechanical dynamics (1) holds interacting
measured quantum object and measurement device [4] etc.

It can be observed that different inputs of given qm-adder
satisfies practically the same quantum mechanical dynamics (1),
i.e. that $\hat {U}_{+}$ is practically independent of the
concrete initial dynamical states of $A$ and $B$. In other words
quantum mechanical dynamics (1) is symmetric in respect to change
of given initial dynamical states of the $A$ and $B$ by some other
from ${\it B}_{A}$ and ${\it B}_{B}$. Now we shall show that this
symmetry stands conserved by more accurate description of the
quantum mechanical dynamical evolution (1) when time dependence of
the quantum states and evolution operator is explicit.

For this reason we shall start from differential form of the
super-systemic dynamical evolution (1) on $A+B$ generalized in
such way that it includes mentioned time dependence
\[(\hat {H}_{A}\otimes \hat {1}+\hat {1}\otimes  \hat {H}_{B} +\hat
{V}_{A} \otimes \hat {V}_{B})|n>|n+m(t)>= \]
\begin {equation}
=(i \hbar \frac {d}{dt} \otimes \hat {1} + \hat {1} \otimes  i
\hbar \frac {d}{dt})|n>|n+m(t)>     for \hspace{0.5cm} \forall n,m
\in W
\end{equation}

Here $\hat {H}_{A}$ and $\hat {H}_{B}$ represents Hamiltonian
observable of the isolated $A$ and $B$, $\hat {1}$ corresponding
unit observable, and $\hat {V}_{A} \otimes \hat {V}_{B}$
observable of the interaction between $A$ and $B$, and $|n+m(t)>$
time dependent dynamical state of the $B$  for $ \forall n,m \in
W$. It will be supposed that $\hat {H}_{A}$ and $\hat {V}_{A}$
commute and that ${\it B}_{A}$ represents their eigen basis. Also,
it will be supposed that spectrum of the eigen values of $\hat
{V}_{A}$ is nondegenerate. Finally, it will be supposed that
initial condition is
\begin {equation}
|n>|n+m(0)>=|n>|m>
\end {equation}
and that $|n>|n+m(t)>$ tends finally to $|n>|n+m>$ for
sufficiently large $t$ for $\forall n, m \in W$. Expression
"sufficiently large" would mean infinite large, but, practically
there is some effective stopping time $T(n,m)$ determined by (2)
and concrete n and m (this determination will not be considered
detachedly here), so that
\begin {equation}
<n+m|<n|n>|n+m(t)> = <n+m|n+m(t)> \simeq 1   \hspace{0.5cm} for
\hspace{0.5cm} t> T(n,m)
\end {equation}
or
\begin {equation}
<n+m+k|<n|n>|n+m(t)> = <n+m+k|n+m(t)> \simeq  0  \hspace{0.5cm}
for \hspace{0.5cm} t > T(n,m)
\end {equation}
for $\forall n,m,k \in W$.

According to introduced suppositions, (2), after partial scalar
product by $<n|$, turns formally  in the sub-systemic quantum
mechanical dynamical evolution on $B$
\begin {equation}
(\hat {H}_{B} + v_{An}\hat {V}_{B})|n+m(t)> = i \hbar \frac
{d}{dt} |n+m(t)>
\end {equation}
with initial condition corresponding to (3), where $ v_{An}$
represents eigen value of  the $\hat {V}_{A}$ in the $|n>$ for
$\forall  n,m \in W$. Obviously, for any concrete $m$ expression
(6) does not represent one dynamical equation but a series of the
different dynamical equations any of which is determined by
corresponding value of the $n$, i.e. by initial dynamical state of
the $A$.

But, according to standard quantum mechanical formalism [4]-[10]
(that is in full agreement with existing experimental data [11],
[12]), it is well-known that super-systemic dynamical evolution
(2) on $A+B$ yields a more complete description of the interaction
between $A$ and $B$ than sub-systemical dynamical evolution (6) on
$B$, even if both dynamical evolutions yield compatible numerical
results. It represents very important fact since it shows that
exact dynamical form of the interaction between $A$ and $B$ is
exactly independent of the initial dynamical states of the $A$ and
$B$. In other word exact quantum mechanical form of the dynamical
interaction between $A$ and $B$ is symmetric in respect to
changing of given initial dynamical states of $A$ and $B$ by some
other from ${\it B}_{A}$ and ${\it B}_{B}$. Or, exact quantum
dynamics of qm-adder is completely independent of the values of
its inputs.

It can be supposed $T(n-k,m)+T(k,m) \geq T(n,m)$ for $\forall n,
m, k \in W$ and $|n| > |k|$ and $|n| > |n-k|$. Then, for $t >
T(n,m)+T(k,m) \geq T(n+k,m)$, according to (2)-(5) it follows that
final dynamical state of $B$ is, practically, $|n+m>$ for $\forall
n,m,k \in W$. However, according to standard quantum mechanical
formalism [4]-[9], i.e. to linear independence of the states from
a (computational) basis, it is satisfied $|n+m> \neq |(n-k)+m> +|
k+m>$  for $\forall  n, m, k \in W$ and $|n|>|k>$ and $|n|>|n-k|$.
It means that quantum mechanical dynamical evolution on $A+B$
(2)-(5) does not admit, even approximately and retrospectively,
its representation by  a succession of the intermediate outputs on
given qm-adder where any of these outputs would be equivalent to
addition of the whole numbers.Practically , it implies, that here
Peano's induction axiom [2],[3] cannot be satisfied, which will be
later (at the end of this section) discussed with more details.

It can be added that in an especial case for $n \geq 0$ and $m
\geq  0$ the same qm-adder realizes the addition of two numbers
$n$ and $m$ from the set of all natural numbers $N$.

Further, according to (1), it follows
\begin {equation}
\hat {U}_{+}|n>|-m>=|n>|n-m>  \hspace{0.5cm} for \hspace{0.5cm}
\forall  n,m \in W
\end {equation}
which means that given qm-adder can to realize arithmetical
operation inverse to addition, the difference, -, of any two whole
number. Also, it means that in an especial case for $n \geq 0$ and
$n \geq m$ the same adder can to realize difference between
natural numbers $n$ and $m$.

Finally, following can be observed. According to standard quantum
mechanical formalism, i.e. mathematical theory of Hilbert's space
of the states (vectors) with unit norm, all states from the
computational basis are linearly independent. In this sense none
of these states has more complex either simpler structure than any
other. In the same sense, all states from computational basis have
the same complexity (simplicity). For example, $|n>$, $|m>$ and
$|n+m>$ have the same complexity (simplicity) in the ${\it B}$, or
$|n>|m>$ and $|n>|n+m>$ have the same complexity (simplicity) in
the ${\it B} \otimes {\it B}$, for $\forall  n, m \in W$. (In
other words  "distance" between arbitrary $|n>$ and $|m>$ from the
computational basis, i.e. $|(<m|-<n|)(|n>-|m>)|^{\frac {1}{2}} =
2^{\frac {1}{2}}$  and it is independent of $n$ and $m$ for
$\forall  n, m \in W$). Finally, according to (1), $\hat {U}_{+}$
acts in the same complex (simple)  way at any state from the ${\it
B}_{A}\otimes {\it B}_{B}$.

All this  points clearly that addition of any two whole (natural)
numbers, realized by qm-adder, has the same complexity
(simplicity) (in previously defined sense) as the addition of any
two other whole (natural) numbers realized by qm-adder. It
represents a principal distinction in respect to addition realized
by corresponding Turing's machine, i.e. to addition within usual,
or, more precisely, G$\ddot {o}$del-Turing-Church's, i.e. GTC
axiomatic system of the arithmetic, or, simply GTC-arithmetic
[1]-[3].

Obviously, both inputs and output of the qm-adder one-to-one
correspond to both inputs and output of Turing's machine for
addition. It means that here Church's thesis is satisfied. (As it
is well-known [1] Church's thesis suggests that {\it any
arithmetical algorithm is equivalent  to corresponding especial
Turing's machine that belongs to universal Turing's machine}. It
can be generalized (which will be discussed later) in the
following way which will be used in the further work. {\it All
inputs and output of any algorithm one-to-one correspond to all
inputs and output of the corresponding especial Turing's machine
that belongs to universal Turing's machine}). But, in
GTC-arithmetic, according to Peano's induction axiom, addition
represents a complex operation reducible in "the immediate
successor of" operation, $s$,as the simplest, i.e. elementary
operation. Namely, since $s(n) = n+1$ , it follows $ n+m = s…s(n)$
(where $s$ is repeated $m$ times), for any two natural numbers $n$
and $m \neq 0$. Equivalently, any natural number $n$ is less
complex than natural number $s…s(n) = m$ (where $s$ is repeated
$m-n$  times), for any natural number  $m > n$.  (In other words,
distance between two natural numbers $m > n$ and $n$ , i.e. $m-n$
depends sharply of the values of these numbers.)

Finally, it is not hard to see that a natural number $n$ in the
GTC-arithmetic can correspond to a vector with norm $n$ from the
one-dimensional vector space of the vectors without constant norm.
Obviously, such one-dimensional vector space of the vectors
without constant norm is conceptually completely opposite to
infinite-dimensional Hilbert's space of the vectors with constant
(unit) norm.

For this reason, as it is not hard to prove, addition realized by
qm-adder cannot be reduced in any simplest operation (here,
obviously, is no analogy with "the immediate successor of "
operation), or it represents really an elementary operation. All
this indicates that qm-adder implies an (axiomatic system of the)
arithmetic principally different from GTC-arithmetic, as it will
be shown and discussed in the further sections of this work.

\section {Quantum multiplier}

Now we shall define  a quantum gate that realizes other well-known
arithmetical operation  , multiplication, denoted by $   $or
simply by blanc symbol, between any two whole or natural numbers
in an effectively  finite time interval. It will be called
qm-multiplier.

Let $A$,$B$ be two quantum systems whose quantum states belong to
${\it B}_{A}$, ${\it B}_{A}$ (all equivalent to ${\it B}$) from
the ${\it H}_{A}$, ${\it H}_{A}$ (all equivalent to ${\it H}$)
respectively.

Let short-lived (which means that here time dependence of
corresponding states and operators will be not given explicitly)
quantum dynamical interaction between A, B or quantum dynamical
evolution on the quantum super-system A+B (that holds both  A and
B representing its sub-systems) be given by unitary evolution
operator $\hat {U}_{\times}$. Let $\hat {U}_{\times}$ satisfies
following conditions
\begin {equation}
\hat {U}_{\times} |n>|m>=|n>|nm> \hspace{0.5cm} for \hspace{0.5cm}
\forall  n, m \in W, n \neq 0
\end {equation}
where $|n>$ and $|m>$ on the left  hand of the (8) represent
initial dynamical states of the $A$ and $B$, while $|n>$ and
$|nm>$ on the right hand of the (8) represent final dynamical
states of the $A$ and $B$ for $\forall  n, m \in W $. Of course
$\hat {U}_{\times}$ acts over ${\it H}_{A}\otimes {\it H}_{B}$.
(For $n=0$ expression (8) must be especially redefined but we
shall not consider this redefinition explicitly. Namely, given
redefinition, without any principal problem, needs increase of the
technical complexity of the qm-multiplier. If further text, for
reason of the simplicity, we shall consider that (8) includes
$n=0$ case too.) Even if $\hat {U}_{\times}$ is not completely
determined by (8) it can be considered that $\hat {U}_{\times}$
represents an unitary evolution operator that satisfies (8).

Now, we shall prove that $A+B$ represents really a  qm-multiplier.

Namely, according to (8) there is one-to-one correspondence
between $A$ initial dynamical state $|n>$ and $B$ initial
dynamical state $|m>$  and multiplication arguments $n$ and $m$
for  n, m $\in$ W. In this way initial dynamical states of A and B
can represent qm-multiplier inputs. Further, according to (8),
there is one-to-one correspondence between  B  final dynamical
state |nm> and  multiplication result nm for  n, m $\in$  W. In
this way final dynamical state of B can represent qm-multiplier
output.

It is obvious that for $\forall  n, m, p, q \in W $ and arbitrary
real numbers $c$, $d$, $e$, $f$ that satisfy conditions
$|c|^{2}+|d|^{2}=1$ and $|e|^{2}+|f|^{2}=1$ from (8) it follows
$\hat {U}_{\times} (c|n>+d|p>)(e|m>+f|q>) =
ce|n>|nm>+cf|n>|q>+de|p>|m>+df|p>|q>$. It means that quantum
multiplier can work by arbitrary superposition of the states from
computational bases, i.e. by "qubits". But such general situation
will not be analyzed in this work.

So, it is proved that described $A+B$ quantum super-system with
quantum mechanical dynamics (8) represents  a qm-multiplier. It
can be added that described qm-multiplier cannot be simply
technically realized, i.e. that $A+B$ must be really a very
complex quantum super-system that includes many sub-systems. In
other words, real qm-multiplier can be only formally, i.e.
effectively  presented to be equivalent to $A+B$. But this fact
does not represent any principal problem for existence of given
$A+B$ qm-multiplier.

On the basis of an analysis, which will not be done explicitly but
that is analogous to analysis from the end of the previous
section, it can be stated following. Even by more accurate
description, when corresponding dynamical states and operators
become time dependent, quantum mechanical dynamics of the
qm-multiplier is symmetric in respect to changing of the values of
qm-multiplier inputs.

Also, multiplication realized by given qm-multiplier is
principally different from multiplication realized by
corresponding Turing's machine, i.e. multiplication within
GTC-arithmetic. Namely, qm-multiplier both inputs and output are
one-to-one corresponding to Turing's machine for multiplication
both inputs and  output which means that Church's thesis is
satisfied. But, multiplication in GTC-arithmetic, represents a
complex operation reductable in the simpler, addition or "the
immediate successor of", i.e. s, operations.(For example, in
GTC-arithmetic, multiplication of two natural number n and m can
be realized by addition of n natural numbers equivalent to m.) On
the other hand, as it is not hard to see, qm-multiplication
represents an operation that has the same complexity (simplicity)
as qm-addition operation. (For example, $\hat {U}_{\times}|n>|m> =
|n>|nm>$ which is different from the $\hat {U}_{+}^{n-1} |m>|m> =
|m>|nm>$, for $n-1,m=1,2, … $.)  In this sense qm-multiplication
represents really an elementary arithmetical operation.

\section {Universal qm-arithmetical gate. Decidable and complete qm-arithmetic}

As it is well-known elementary logical operations can be defined
by arithmetical operations in following way
\begin {equation}
\neg p =1-p   \hspace{1.5cm}  (negation, NO)
\end {equation}
\begin {equation}
p \wedge q = pq  \hspace{1.5cm} (conjunction, AND)
\end {equation}
\begin {equation}
p \vee q = p+q-pq  \hspace{1.5cm} (disjunction, OR)
\end {equation}
etc., for $\forall p,q$ that belong to ${0,1}$. Here 0 corresponds
to logical untruth while 1 corresponds to logical truth. All other
composed logical operations according to a logical recursions,
i.e. induction rules, can be obtained by an induction by previous
elementary  logical operations. Roughly speaking usual
(propositional)  logic represents an especial sub-theory of the
arithmetic.

Now we shall determine all possible quantum gates representing
physical models of the algorithms, precisely qm-algorithms for
realization of corresponding arithmetical (including logical)
operations, denoted qm-arithmetical gates, that satisfy following
conditions. They work in some effectively finite time intervals
and they can be defined recursively, i.e. by inductive combination
of the  elementary qm-arithmetical gates. Also, it means that
given qm-arithmetical gates satisfies Church's thesis.

Ordered series of all such qm-arithmetical   gates will be simply
called universal qm-arithmetical gate. Also, axiomatic system of
the arithmetic based on the universal qm-arithmetical  gate will
be called qm-arithmetic and its operations - qm-arithmetical
operations.

For reason of simplicity in further work we shall not differ
explicitly a qm-arithmetical gate and corresponding qm-algorithm
for realization of corresponding qm-arithmetical operation. In
further simplification we shall not differ explicitly qm-algorithm
and corresponding qm-arithmetical operation if this operation can
be realized by given algorithm. In this sense, for arbitrary
natural number $n$, we shall not differ explicitly this number and
quantum state $|n>$.

We shall consider that some qm-arithmetical operation is
elementary if it cannot be reduced in some other qm-arithmetical
operations and if it can be realized (by a qm-arithmetical
algorithm) in the completely same way (in the completely same
number of the algorithm steps) for any natural numbers that
represent its arguments.

Let $M_{0}(n)$ be unary qm-arithmetical operation (free variable
operation) that applied on any natural number $n$ determines given
number as the free variable, i.e. that forbids that this number be
result of any qm-arithmetical operation. Obviously, given
operation does not represent result of any other qm-arithmetical
operation. Also, it is satisfied $ M_{0}(n) = n$  for any natural
number n which means that given operation is realized in the
completely same way for any natural number so that there is no
need  that its argument value be given explicitly . For this
reason given operation can be simply denoted  $M_{0}$. Thus,
$M_{0}$ represents  an elementary unary qm-arithmetical operation.

Suppose that  $M_{0}$ represents the unique elementary unary
qm-arithmetical operation. (This supposition can be
formally-mathematically considered as an axiom.)

Let $M_{1}(n, k)$ be binary  qm-addition qm-arithmetical operation
that applied on any natural number $n$ as its first argument and
any natural number $k$ as its second argument both representing
the free variables yields the result  $n+k$. Suppose that
qm-addition represents an elementary binary qm-arithmetical
operation. Elementarily of the qm-addition follows from the
definition of the qm-adder, i.e. from the unitary symmetric
quantum mechanical dynamics. (But, this elementarily can be
formally-mathematically considered as an axiom.) In accordance
with previous discussions and suppositions  instead of the
$M_{1}(n,k)$ we can write $M_{1}(M_{0}(n), M_{0}(k))$ or
$M_{1}(M_{0}, M_{0})$ or only $M_{1}$.

Further, let  $M_{2}(n,k)$ be binary qm-multiplication, i.e. such
qm-arithmetical operation that applied on any natural number $n$
as its first argument and any natural number $k$ as its second
argument both representing the free variables yields the  result
$nk$. Suppose that qm-multiplication represents an elementary
binary qm-operation. Elementarity of the qm-multiplication follows
directly from the definition of the qm-multiplier, i.e. from the
unitary symmetric quantum mechanical dynamics.(But this
elementarity can be formally-mathematically considered as an
axiom). In accordance with previous discussions and suppositions
instead of $M_{2}(n,k)$ we can write $M_{2}(M_{0}(n),M_{0}(k))$ or
$M_{2}(M_{0},M_{0})$ or only $M_{2}$.

Suppose that except $M_{1}$ and $M_{2}$ other elementary binary
qm-arithmetical operations do not exist. This supposition follows
directly from the characteristics of the unitary symmetry of  the
quantum mechanical dynamics. (But, this supposition can be
formally-mathematically considered as an axiom). Suppose that all
other qm-arithmetical operations can be obtained by corresponding
rules, i.e. induction starting form the $M_{0}$, $M_{1}$ and
$M_{2}$, so that these obtained qm-arithmetical operations are not
elementary. Given induction can be realized in the following way.

Firstly, we shall define qm-arithmetical operations
\begin {equation}
M_{i}(M_{j},M_{k})  \hspace{0.5cm} for \hspace{0.5cm}  i=1, 2 ;
j=0,1,2 ; k=0,1,2 ; j \neq k          .
\end {equation}
In fact  (12) denotes qm-arithmetical operations obtained by
application of the  qm-arithmetical operation  $M_{i}$ on the
qm-arithmetical operation $M_{j}$ as its first and qm-arithmetical
operation  $M_{k}$ as its second argument, under conditions $
i=1,2 ; j=0,1,2 ; k=0,1,2  ; j \neq k $. Obviously, all given
qm-arithmetical operations (12) are well-defined.

Qm-arithmetical operations (12) can be unambiguously enumerated,
using (including restriction  conditions) lexicographic rules for
notation of the variations with repetitions of the elements
$M_{0}$,$M_{1}$ and $M_{2}$ of the third class, in the following
way
\begin {equation}
M_{3}=M_{1}(M_{0},M_{1})  \Leftrightarrow  n + (m+k)
\end {equation}
\begin {equation}
M_{4}=M_{1}(M_{0},M_{2})  \Leftrightarrow   n + (m\cdot k)
\end {equation}
\begin {equation}
M_{5}=M_{1}(M_{1},M_{0})  \Leftrightarrow   (n+m) + k
\end {equation}
\begin {equation}
M_{6}=M_{1}(M_{1},M_{1}) \Leftrightarrow   (n+m)+(k+l)
\end {equation}
\begin {equation}
M_{7}=M_{1}(M_{1},M_{2}) \Leftrightarrow (n+m)+(kl)
\end {equation}
\begin {equation}
M_{8}=M_{1}(M_{2},M_{0})   \Leftrightarrow  (n\cdot m)+k
\end {equation}
\begin {equation}
M_{9}=M_{1}(M_{2},M_{1})  \Leftrightarrow  (n\cdot m)+(k+l)
\end {equation}
\begin {equation}
M_{10}=M_{1}(M_{2},M_{2})   \Leftrightarrow   (n\cdot m)+(kl)
\end {equation}
\begin {equation}
M_{11}=M_{2}(M_{0},M_{1})  \Leftrightarrow    n(m+k)
\end {equation}
\begin {equation}
M_{12}=M_{2}(M_{0},M_{2})  \Leftrightarrow   n(m\cdot k)
\end {equation}
\begin {equation}
M_{13}=M_{2}(M_{1},M_{0})  \Leftrightarrow  (n+m)k
\end {equation}
\begin {equation}
M_{14}=M_{2}(M_{1},M_{1})  \Leftrightarrow  (n+m)(k+l)
\end {equation}
\begin {equation}
M_{15}=M_{2}(M_{1},M_{2})  \Leftrightarrow  (n+m)(kl)
\end {equation}
\begin {equation}
M_{16}=M_{2}(M_{2},M_{0})   \Leftrightarrow  (n\cdot m)k
\end {equation}
\begin {equation}
M_{17}=M_{2}(M_{2},M_{1})   \Leftrightarrow  (n\cdot m)(k+l)
\end {equation}
\begin {equation}
M_{18}=M_{2}(M_{2},M_{2})  \Leftrightarrow  (n\cdot m)(kl)
\end {equation}
Here, on the right-hand sides of  $\Leftrightarrow $ explicit
forms of corresponding qm-arithmetical operations are given where
natural numbers $n$, $m$, $k$, $l$ represent corresponding
arguments of $M_{0}$, $M_{1}$, $M_{2}$ and where small brackets
denote results of the qm-addition or qm-multiplication.

It is not hard to see that final results of some individual
qm-arithmetical operations are mutually equivalent, eg. $M_{3}$
(13) and $M_{5}$ (15). However, since qm-addition and
qm-multiplication are elementarity and mutually  different, given
individual qm-arithmetical operations, eg. $M_{1}(M_{0}, M_{1})$
(13) and $M_{1}(M_{1},M_{0})$ (15), are different too.

Thus,(13)-(28) define all composite qm-arithmetical operations
that can be defined inductively, by one elementary qm-arithmetical
operation whose one or both arguments,representing completely free
variables,are changed by variables that must have form of some of
the elementary qm-arithmetical operations.

Obviously, an ordered enumerable series of all composite
qm-arithmetical operations, i.e. universal qm-gate can be obtained
by further induction. Concrete technical realization of the
universal qm-gate, that unambiguously exists, will not be
considered explicitly here.

It is not hard to see too that in any finite step of given
induction corresponding composite qm-arithmetical operations, i.e.
qm-arithmetical gates  work in some effectively finite time
intervals. In this sense universal qm-arithmetical operation, i.e.
universal qm-gate work in the effectively finite time intervals.

Here, obviously  (which will not be proved explicitly), subscript
of any composite qm-arithmetical operation  is determined
unambiguously and it corresponds, practically (i.e. including
small corrections), to lexicographic number of the variations of
three elements, $M_{0}$,$M_{1}$ and $M_{2}$, of corresponding
finite class,  with repetitions.

Also, it is not hard to see that all (elementary or composite)
qm-arithmetical operations can be divided in the disjunctive
classes in the following way.

Zeroth class holds  all elementary qm-arithmetical operations
$M_{0}$, $M_{1}$, $M_{2}$  whose all, one or two, arguments
represent  completely free variables.

First class holds all elementary qm-arithmetical operations whose
at least one argument represents an elementary qm-arithmetical
operation from the zeroth class. …

Generally, $n$-th class holds all elementary qm-arithmetical
operations whose at least one argument represents an elementary
arithmetical operations from $(n-1)$-th class.

In this way all classes of  the qm-arithmetical operations are
defined recursively, i.e. inductively.

According to introduced definition of the classes qm-arithmetical
operations can be expressed by
\begin {equation}
M^{k}_{\alpha}=M^{0}_{i}(M_{0},M^{k-1}_{j})
\end {equation}
\begin {equation}
M^{k}_{\beta}=M^{0}_{i}(M^{k-1}_{j},M_{0})
\end {equation}
\begin {equation}
M^{k}_{\gamma} =M^{0}_{i}(M^{k-1}_{j},M^{k-1}_{m})
\end {equation}
Here $M^{0}_{i}=M_{i}$ for $i=0,1,2$  represents an elementary
qm-arithmetical operations, i.e. qm-arithmetical operations from
the zeroth class ; $M^{k-1}_{j}$  -  qm-arithmetical operations
from the $(k-1)$-th class ; and
$M^{k}_{\alpha}$,$M^{k}_{\beta}$,$M^{k}_{\gamma}$   -
qm-arithmetical operations from the $k$-th class

As it is not hard to see (which will not be proved explicitly),
subscripts
\begin {equation}
\alpha = \alpha (i,0,j,k)
\end {equation}
\begin {equation}
\beta = \beta (i,j,0,k)
\end {equation}
\begin {equation}
\gamma = \gamma  (i,j,m,k)
\end {equation}
represent uniquely determined functions of theirs arguments, i.e.
subscripts 0, $i$, $j$, $m$, $k$, more precisely there are
one-to-one correspondences between $\alpha$, $\beta$, $\gamma$
functions and theirs arguments 0, $i$, $j$, $m$, $k$.

Formally, (29)-(31)  can be presented in the following generalized
form
\begin {equation}
M^{k}_{\delta}=M^{0}_{i}({\it a},{\it b}) \hspace{0.5cm} for
\hspace{0.5cm}   i =1,2
\end {equation}
where  either ${\it a} = M_{0}$ and ${\it b}$ represent
qm-arithmetical operation from the $(k-1)$-th class, either ${\it
a}$ represents qm-arithmetical operation from the $(k-1)$-th class
and ${\it b} = M_{0}$, and where
\begin {equation}
\delta = \delta (i,{\it a}, {\it b})
\end {equation}
represents  a formal generalization of (32)-(34).

According to previous discussions and (35), (36) left hand  of
(35), $ M^{k}_{\delta (i, {\it a}, {\it b})}$, represents uniquely
determined, $\delta (i, {\it a}, {\it b})$-th in the ordered
series of the qm-arithmetical operations (from $k$-th class).

Also, according to previous discussions and (35), (36), right-hand
side of (35), $M^{0}_{i}({\it a}, {\it b})$, represents uniquely
determined zeroth class qm-arithmetical operation, that acting at
$ ({\it a}, {\it b})$  as its argument, yields $M^{k}_{\delta
(i,{\it a}, {\it b})}$ as its result.

Finally, it is very important to be pointed out that {\it
according to previous discussions and} (35), (36) {\it for given}
$k$, {\it there is no  one-to-one correspondence between
subscript} $i$ {\it and} $ ({\it a},{\it b})$  {\it on the one
hand , and that subscript} $\delta $ {\it one-to-one corresponds
to} $(i, {\it a}, {\it b})$ {\it in} (35), (36) {\it on the other
hand}.

Ordered series of the qm-arithmetical operations (35),(36) for any
natural number $k$  represents universal qm-arithmetical
operation.

It is not hard to see (which will not be proved explicitly) that
final  result of any qm-arithmetical operation realized (decided)
by universal qm-arithmetical gate one-to-one corresponds to final
results of the same qm-arithmetical operations realized (decided)
by universal Turing's machine and vice versa. In this way Church's
thesis is satisfied.

Now we shall prove that in the qm-arithmetic there are none
theorem analogous to Church's undecidability theorem [3] or
G$\ddot {o}$del's incompleteness theorem [1], [2] existing in the
GTC-arithmetic. I.e. we shall prove that qm-arithmetic, in
distinction to GTC-arithmetic, is decidable and complete.

As it is well-known [3], Church's undecidability theorem that
implies G$\ddot {o}$del's incompleteness theorem, can be
formulated in the GTC-arithmetic in the following way. In the
GTC-arithmetic there is a set $S_{A}$ of the arithmetical
formulas. A formula from the $S_{A}$ can be denoted by  $x$. Also,
in the GTC arithmetic there is an enumerable ordered series of the
algorithms simply denoted by $k$ for $k=1,2, … $ so that $k \in
N$. A decidable arithmetical function  or operation in the GTC
arithmetic $f_{k}(x)$ represents the result of the action of some
arithmetical algorithm $k$ at any formula $x$. But Church's
function $f(x)=f_{x}(x)+1$ is undecidable. Namely, supposition
that $f(x)$ is decidable, i.e. that there is such m for which
$f(x)=f_{m}(x)$, yields, by G$\ddot {o}$del-Church's
digitalization procedure, i.e. for $x=m$, a contradiction
$f_{m}(m) = f_{m}(m)+1$. In this way Church's undecidability
theorem, or theorem of the existence of undecidable formulas in
GTC-arithmetic is proved.

It is very important that following be pointed out. By definition
of a decidable arithmetical function or operation  $f_{k}(x)$, $k$
from the $N$ and $x$ from the $S_{A}$ represent mutually
independent variables. Then, for some fixed $k$ from $N$, $x$ can
hold all possible values from $S_{A}$.Without this fact G$\ddot
{o}$del-Church's diagonalization  and Church's undecidability
theorem cannot be formulated.

Now suppose that $f_{k}(x)$ in the GTC-arithmetic can one-to-one
correspond to $M^{k}_{\delta}= M^{0}_{i}({\it a},{\it b}) $
(35),(36) in the qm-arithmetic. Then  $k$ from the  $f_{k}(x)$
must correspond to $(k,\delta)$ from the $M^{k}_{\delta}$  and,
simultaneously, $x$ from the  $f_{k}(x)$ must correspond to $(i
,{\it a},{\it b})$  argument of the $\delta$ . But, as it has been
discussed, $k$ and $x$ represent mutually independent variables
while $\delta$ and $(i,{\it a},{\it b})$  are strictly dependent
since there is one-to-one correspondence between $\delta$ and
$(i,{\it a},{\it b})$ . In this way it is shown that previous
supposition is incorrect so that there is none unambiguous
correspondence between decidable functions or operations in the
GTC-arithmetic and qm-arithmetic.

For this reason in the qm-arithmetic a diagonalization procedure
analogous to diagonalization procedure in the GTC-arithmetic
cannot exist. Moreover, for the same reason, in the qm-arithmetic
a theorem analogous to Church's undecidability theorem or G$\ddot
{o}$del's incompleteness theorem in the GTC-arithmetic cannot
exist.

In this way it  is proved  that given qm-arithmetic is decidable
and complete.

\section {Discussion and conclusion}

As it is well-known Feynman [13], Deutsch [14] and some other
scientists suggested that quantum computers, i.e. computers whose
working is based on the quantum mechanical dynamics (including its
characteristic superposition principle) [15], can be more
efficient in the practice than any classical computer (including
universal Turing's machine), i.e. computer whose working is based
on the classical mechanical dynamics. Really, Shor (Shor's
factorization ) [16] and Grover (Grover's algorithm for data base
search) [17] showed that there are such quantum algorithms
realizable by quantum computers that are faster than any classical
algorithm realizable by classical computers. Moreover, even if
Deutsch [14] proved that classical Church's thesis [1] can be
simply (almost "trivially") statistical generalized to be quantum
Church's thesis, Grover [18] pointed out : "The quantum search
algorithm is a technique for searching N possibilities in only O(
N) steps. Although the algorithm itself is widely known ,not so
well known is the series of the steps that first lead to it, these
are quite different from any of the generally known forms of the
algorithm." In other words Grover's quantum algorithm for data
base search  is not only faster, but completely different from any
Turing's algorithm. All this opens many serious questions not only
on the correlations between classical and quantum algorithms but
also between (foundation of the) mathematics (arithmetic) and
(foundation of the) physics (quantum mechanics). For example,
Deutsch [14] states that Church's thesis expresses, in fact, a
physical principle, etc..Now we shall discuss some of these
questions  and we shall suggest some possible answers.

First of all, as it has been presented in the section 2., we use a
generalized form of the usual Church's thesis [1]. Namely, we
suggest, that generalized Church's thesis needs one-to-one
correspondence  between all inputs and output of any algorithm and
corresponding especial Turing's machine only. Simultaneously we
suggest that this thesis does not need necessarily one-to-one
correspondence  between functional dependence between inputs and
output of given algorithm on the one, and, functional dependence
between inputs and output of the corresponding especial Turing's
machine on the other hand. Thus, generalized Church's thesis can
be considered as a consistent general definition of any
representation of the decidable part of the arithmetic. Also, in
contrast to usual Church's thesis, it admits significantly larger
number of the especial representations of the decidable part of
arithmetic.

Generalized Church's thesis admits GTC-arithmetic [1]-[3] as an
especial case.In this GTC-arithmetic, that includes usual Church's
thesis, practically all especial representations of its decidable
part are equivalent to  mathematical structure of a discrete
one-dimensional vector space of the vectors without constant norm
. In this space dominant forms of the "motion" are discrete (for
natural and whole numbers) translations defined, in fact, by
Peano's  induction axiom [2],[3]. But such GTC-arithmetic ,
according to Church's undecidability theorem [1] and G$\ddot
{o}$del's incompleteness theorem [2],[3], admits existence of the
undecidable formulas (operations),i.e. its undecidable parts. In
this sense GTC-arithmetic is incomplete.

However, as it has been proved, generalized Church's thesis admits
that decidable part of the arithmetic, i.e. qm-arithmetic, be
presented in an oposit way, i.e. by mathematical structure of the
infinite-dimensional Hilbert's space of the vectors with constant
(unit) norm. In this space dominant forms of the "motion" are
discrete unitary transformations, i.e. "rotations" (accurate form
of the superposition principle!) that one-to-one map computational
basis in the same computational basis. (Quite naturally and simply
these "motions" can be generalized by such unitary transformations
that one-to-one map computational basis in any other basis in
given Hilbert's space. In this case superposition principle and
existence of the qubits become explicit.) According to well-known
characteristics of such "motions" (norm definition in given
Hilbert's space) here Peano's induction axiom, Church's
undecidability theorem and G$\ddot {o}$del's incompleteness
theorem cannot be consistently defined. For this reason
qm-arithmetic has no undecidable part, or, it is decidable and
complete which admits that some principal open problems in the
foundation of the arithmetic [1]-[3] can be consistently solved.

Finally, it is very important to be pointed out that mathematical
characteristics of the qm-arithmetic are, in fact, completely
independent of the physical characteristics of the quantum systems
representing qm-arithmetical gates. Or, simply speaking, in
distinction to Deutsch opinion [14], Church's thesis does not
represent any physical principle. Namely, qm-arithmetic as a
mathematical theory is practically completely determined by
generalized Church's thesis and mathematical theory of the
infinite-dimensional Hilbert's space of the unit norm vectors.
But, since standard quantum mechanical formalism uses the same
mathematical theory of the Hilbert's space, qm-arithmetic can be
consistently physically modeled by quantum mechanics.

\section {References}

\begin {itemize}

\item [ [1] ] S.C.Kleene, {\it  Mathematical Logic} (John Wiley and Sons, New York,1967.)
\item [ [2] ] K.Gödel, {\it On Formally Undecidable Propositions of Principia Mathematica and Related Systems} (Dover, New York, 1992.)
\item [ [3] ] E.Nagel, J.R.Newman, {\it G$\ddot {o}$del's Proof} (New York University Press, New York, 2001.)
\item [ [4] ] J.von Neumann, {\it Mathematische Grundlagen der Quanten Mechanik} (Springer Verlag, Berlin, 1932.)
\item [ [5] ] P.A.M.Dirac, {\it Principles of Quantum Mechanics} (Clarendon Press, Oxford, 1958.)
\item [ [6] ] A.Messiah, {\it Quantum Mechanics} (North-Holand Publ.Co., Amsterdam,1970.)
\item [ [7] ] B.d'Espagnat, {\it Conceptual Foundations of Quantum Mechanics} (Benjamin, New York, 1976.)
\item [ [8] ] N.Bohr, {\it Atomic Physics and Human Knowledge} (John Wiley and Sons, New York, 1958.)
\item [ [9] ] N.Bohr, Phys.Rev., {\bf  48 }, (1935.), 696.
\item [ [10] ] J.S.Bell, Physics, {\bf  1}, (1964.), 195.
\item [ [11] ] A.Aspect, P.Grangier, G.Roger, Phys.Rev.Lett., {\bf  47}, (1981.), 460.
\item [ [12] ] A.Aspect, J.Dalibard, G.Roger, Phys.Rev.Lett., {\bf  49}, (1982.), 1804.
\item [ [13] ] R.Feynman, Found.Phys., {\bf  16}, (1986.), 507.
\item [ [14] ] D.Deutsch, Proc.R.Soc.London A, {\bf  400 }, (1985.), 97.
\item [ [15] ] M.A.Nielsen, I.L.Chuang, {\it Quantum Computation and Quantum Information} (Cambridge University Press, Cambridge, 2000.)
\item [ [16] ] P.Shor, SIAM Journal of Computing, {\bf  26}, (1997.), 1484.
\item [ [17] ] L.K.Grover, Phys.Rev.Lett., {\bf  79}, (1997.), 325.
\item [ [18] ] L.K.Grover, {\it From Schrödinger's Equation to the Quantum Search Algorithm}, quant-ph/0109116.

\end{itemize}

\end{document}